\documentclass[12pt]{article}

\usepackage{graphicx}
\usepackage{fullpage,amsmath,amssymb,latexsym}
\usepackage{multirow}
\usepackage{natbib}

\begin{document}
\title{Two Empirical Regimes of the Planetary Mass-Radius Relation}





\author{Dolev Bashi$^1$\footnote{corresponding author: dolevbashi@gmail.com} , Ravit Helled$^{1,2}$, Shay Zucker$^1$ and Christoph Mordasini$^3$\\
{\small $^1$School of Geosciences, Tel-Aviv University, Tel-Aviv, Israel}\\
{\small $^2$ Center for Theoretical Astrophysics \& Cosmology,} \\
{\small Institute for Computational Science, University of Zurich, Switzerland.}\\
{\small $^3$Physics Institute, University of Bern, Switzerland.}
 }
  
\date{}
\maketitle
  
\begin{abstract}
Today, with the large number of detected exoplanets and improved
measurements, we can reach the next step of planetary
characterization. Classifying different populations of planets is not
only important for our understanding of the demographics of various
planetary types in the galaxy, but also for our understanding of
planet formation. We explore the nature of two regimes in the
planetary mass-radius (M-R) relation. We suggest that the transition
between the two regimes of "small" and "large" planets, occurs at a
mass of $124 \pm 7$, $M_\oplus$ and a radius of $12.1 \pm 0.5$,
$R_\oplus$. Furthermore, the M-R relation is $R \propto M^{0.55\pm
0.02}$ and $R \propto M^{0.01\pm0.02}$ for small and large planets,
respectively. We suggest that the location of the breakpoint is linked
to the onset of electron degeneracy in hydrogen, and therefore,
to the planetary bulk composition. Specifically, it is the
characteristic minimal mass of a planet which consists of mostly
hydrogen and helium, and therefore its M-R relation is determined by
the equation of state of these materials. We compare the M-R
relation from observational data with the one derived by
population synthesis calculations and show that there is a good
qualitative agreement between the two samples.  
\end{abstract}

%

\section{Introduction}
Exoplanet studies have now reached the level at which planet
characterization is possible. There are hundreds of planets with
measured masses and radii. Knowledge of these two physical properties
provides valuable clues about the planetary composition, through the
mass-radius (hereafter M-R) relationship.  Traditionally, planets have
been divided into two main groups. The first includes the massive,
gas-dominated planets, while the second consists of the small,
terrestrial planets (e.g., Weidenschilling 1977). In part, this
division is inspired by the Solar System, where massive planets are
composed of volatile materials (e.g., Jupiter) while the terrestrial
planets are small and consist of refractory materials. However, the
diversity in masses and radii of exoplanets\footnote{See {\it
http://exoplanets.org} for exoplanet properties.} has taught us that
this separation is somewhat arbitrary and may be over-simplistic (see
review by Baraffe et al.\ 2014 and references therein).
\par

While the first detected exoplanets had relatively large masses and
radii, in the recent few years the number of small exoplanets
increased dramatically, due to improvements in technology and
detections from space (e.g., CoRoT (Baglin et al.\ 2006) and
Kepler (Borucki et al.\ 2010)). Of course, since most exoplanets have
been detected via radial velocity measurements or transits, there is a
difference when defining a "small planet" by mass or by radius. In
terms of mass, it is customary to define small planets as planets
whose masses are less than $\sim 30 M_\oplus$ (Mayor et al.\ 2011,
Howard et al.\ 2010), while in terms of radius, small exoplanets are
often those whose radii are smaller than $4 R_\oplus$ (e.g.,
Marcy et al.\ 2014, Weiss \& Marcy 2014). These divisions are
partially based on the behavior of the planetary mass function of
exoplanets.
\par

Previous studies that examined the M-R relation have suggested a
transition in the M-R relation between 'small' planets (Neptune-like)
and 'large' planets (Jovian). Based on visual estimates of the M-R and
mass-density relations, Weiss et al.\ (2013) suggested that the
transition point occurs at a mass of $\sim 150 M_\oplus$. The derived
slopes of the M-R relations in the different regimes turned out to be
$R \propto M^{0.54}$ for $M_\mathrm{p} < 150 M_\oplus$ and $R
\propto M^{-0.039}$ for massive planets ($M_\mathrm{p} > 150 M_\oplus$).
Hatzes \& Rauer (2015) have analyzed changes in the slope of the
mass-density relation. Using a similar slope criterion, they locate
the breakpoint at a mass of $\sim 0.3 M_{J} \simeq 95 M_\oplus$. In a
recent study, Chen \& Kipping (2017) presented a detailed forecasting
model built upon a probabilistic M-R relation using MCMC. According to
their classification the transition between small and large planets
occurs at $0.41 \pm 0.07 M_{J} \simeq 130 \pm 22 M_\oplus$,
corresponding to the transition between Neptunians and Jovians, with
slopes of $R \propto  M^{0.59}$ and $R \propto  M^{-0.04}$ for the low-mass
and high-mass planets, respectively. Interestingly, although the
studies do not agree exactly on the transition mass between the two
regimes, they do agree that it is significantly higher than the
traditional cutoff at $20-30 M_\oplus$. This essentially suggests that
the change in the occurrence rate as seen in the mass function of
exoplanets (at $\sim 30 M_{\oplus}$), i.e., the frequency of planets
is not the same as the behavior of the M-R relation, which is linked
to the planetary composition.

In this paper we present the results of a study we performed in
order to empirically characterize the transition point between small and large planets based on their M-R
relation.  
On the one
hand our aim was to perform a quantitative straightforward study, that
will come up with simple numerical information -- the two M-R
power law indices, and the transition mass. On the other hand, we
opted for a kind of least-square fit, and not an elaborate
probabilistic recipe. Our hope was that this would allow a more
intuitive yet rigorous characterization of the planetary M-R relation.
Finally, we also compare the exoplanet population to formation models and find a qualitative good agreement.  

\section{Sample}

The data we use include only planets with masses and radii that are
based directly on the observations, as opposed to being inferred from
planetary physics models.  Our sample consists of $274$ exoplanets
 queried from {\it http://exoplanets.org} on March 2016. The
lowest mass planet in our sample is Kepler-138b, with a mass of
$0.0667 \pm 0.0604 M_{\oplus}$, well below Earth mass. The highest
planet mass in our sample is that of CoRoT-3b, with a mass of $6945
\pm 315 M_\oplus$($= 21.85 \pm 0.99 M_J$) -- a brown dwarf.  For all
the planets our sample must include measured masses, radii, and their
uncertainties. Thus, we exclude planets with reported masses estimated
based on a theoretical M-R relation. All planets in the sample
are transiting planets, whose masses have been measured either by RV
($238$ through RV follow-up, and $9$ were first detected by RV), or
using TTVs (e.g. Nesvorn\'{y} \& Morbidelli 2008; $27$
planets)\footnote{It should be noted that almost all the TTV planets
are of low mass}. The top panel of Fig.~1 shows the resulting M-R
diagram\footnote{The list of the planets we use is summarized in Table
2 in the online version.}.
\par

\section{Analysis}

The model we assume is that of two mass regimes that differ by the M-R
power law. In the log-log plane, this translates into a continuous
piecewise linear function, with two segments, that we had to fit to
the data points.  In spite of our ambition to apply the most 
basic techniques of simple regression to perform this fit, several
problems conspire to turn this into a somewhat more complicated
problem.

First, the two variables -- the planetary mass and radius -- are both
measured with non-negligible errors. If we could assume that only one
of them (e.g. the mass) had errors, the problem could have been
treated as a standard regression problem. The fact that uncertainties
exist for both variables, takes us to the field of
'errors-in-variables' (EIV) problems, which are surprisingly more
difficult than standard regression problems (e.g. Durbin 1954), and
there is not one agreed approach to analyze them.

As difficult as EIV problems are, in our case the complexity is even
exacerbated by the fact that we aim to fit not a linear function, but
a continuous piecewise-linear function, rendering futile any hope to
solve the problem analytically. Even under the assumptions of standard
regression, where the so-called explanatory variable has no
uncertainty, the problem (dubbed 'segmented regression') is not
trivial (e.g. Hinkley 1969).

Another difficulty arises because of the nature of our specific
sample.  A glance at the top panel of Figure 1 reveals that the data
points are not scattered evenly across the logarithmic mass range. The
points corresponding to the smaller planets seem to be much more
sparse than those of the Jovian planets. The same is true for the very
large planets, with masses of a few Jupiter's mass. One may say that
there seem to be three mass intervals with varying density of
sample points. The origin of this differentiation lies beyond
the scope of this study, and in any case it may very well be a
combination of observational bias and astrophysical processes of
formation and evolution. The smaller number of massive planets
(above a Jupiter-mass) is a result of the low occurrence rate of such
planets (e.g., Cumming et al.\ 2008), while the clustering of small is
probably a result of the massive efforts to detect low-mass planets
and their high occurrence rate (e.g., Howard et al.\ 2012). 

There are various reasons for this sampling variability, ranging from
observational biases to physical effects related to planet formation
and evolution. However, the fact remains that for the purposes of
regression analysis, the mass affects the sampling. In regression
theory this amounts to 'endogenous sampling'.  While fitting a simple
straight line might be affected only slightly by this imbalanced
sampling, it is not guaranteed for a piecewise-linear function. Any
fitting procedure should take this imbalance into consideration.

To streamline the discussion, let us denote:
\begin{eqnarray}
x = \log M_\mathrm{p}\\
y = \log R_\mathrm{p}
\end{eqnarray}
The choice of logarithm base is irrelevant as long as it
consistent throughout the calculation. Eventually the values in linear
scale are the important ones, not in logarithm scale. Now our sample,
in the log-log plane, consists of a set of ordered pairs
$(x_i,y_i)$. Let us further denote by $\Delta x_i$ and $\Delta y_i$
the corresponding logarithmic uncertainties derived from the
uncertainties in the linear scale using the standard transformation.
In cases where the transformation led to assymetric uncertainties, we
still assigneed symmetric errors, by taking the more conservative
(larger error) of the two error estimates.

In our quest for the best-fit piecewise-linear function, we chose what
is probably the most intuitive approach to EIV: Total Least Squares --
TLS (e.g. Markovsky et al.\ 2010). Similarly to standard regression, in
TLS the problem is represented as a minimization problem of a sum of
squares. Each data point contributes to the total sum-of-squares its
orthogonal distance from the fitted line, measured in units of the two
uncertainties. In the simple case where we fit a simple linear
function, if we denote the slope and intercept of the line by $a$ and
$b$, the contribution of the point $(x_i,y_i)$ would be:
$$
\frac{(y_i - a x_i - b)^2}{a^2(\Delta x_i)^2 + (\Delta y_i)^2} \ ,
$$ 
where $\Delta x_i$ and $\Delta y_i$ are the errors of $x_i$ and
$y_i$.  Golub (1973), and Golub \& Van Loan (1980) were the first to
introduce an algorithm to solve this basic TLS problem, using singular
value decomposition. They have also shown that even in this simple
linear case a solution is not guaranteed to exist.

In our case, where the function we seek consists of two straight
lines, we simply calculate, for each point, its weighted orthogonal
distances from the two lines, and include the smaller one in the total
sum-of-squares:
$$ 
S(a_1,b_1,a_2,b_2) = \sum_{i=1}^N \mathrm{min}
\Big\{ \frac{(y_i - a_1 x_i - b_1)^2}{a_1^2(\Delta x_i)^2 + (\Delta
y_i)^2} \ , \frac{(y_i - a_2 x_i - b_2)^2}{a_2^2(\Delta x_i)^2 +
(\Delta y_i)^2} \Big\} \ \ , 
$$ 
where {$N$ is the total number of points and} $a_1$,$b_1$,$a_2$ and
$b_2$ are the slopes and intercepts of the two straight lines. $S$ is
parameterized by four numbers whose meaning is somewhat
arbitrary. This is true especially for the two intercepts $b_1$ and
$b_2$, which are functions of the arbitrary location of the zero point
of $x$. We can instead parameterize $S$ by an alternative,
 physically more meaningful, quadruple: the two coordinates of the
breakpoint (breakpoint mass and corresponding radius), and the two
slopes of the separate mass regimes.

When we set out to minimize $S$, it turned out that the solution was
numerically unstable. Using diffferent starting points for the
optimization algorithm (Nelder-Mead simplex algorithm, see Nelder
\& Mead 1965) resulted in different solutions. That meant that around
the global minimum of $S(a_1,b_1,a_2,b_2)$ there were many local
minima. We suspect that this instability resulted from the endogenous
sampling problem to which we alluded above (the mass-
distribution shown in Fig.~2). In order to rectify this problem we
have decided to introduce weights to the definition of $S$, that will
balance the effect each mass range has on the final solution. As
is clear from the top panel of Figure 1, there are apparently three intervals: $M_\mathrm{p}<69 M_\oplus$, $69 M_\oplus \le
M_\mathrm{p} < 1660 M_\oplus$, and $1660 M_\oplus \le
M_\mathrm{p}$\footnote{We have decided it is beyond the scope of this
study to perform a rigorous clustering analysis. There seems to be a
consensus in data-mining literature that at this stage there is not a
single clustering algorithm or criterion that is guaranteed to be the
best one.  Thus, an intuitive division at this stage is completely
acceptable (e.g., Estivill-Castro 2002).}. Fig.~2 further demonstrates
the differentiation in mass by portraying a histogram of the
mass, together with the borders we chose among the three mass
ranges.

The weighting scheme we applied is known in statistics as Inverse
Probability Weighting, which is designed in order to alleviate the
implications of endogenous sampling (e.g., Wooldridge 1999).  We thus
multiplied the contribution of each data point by a weight that was
supposed to compensate for the effect of the size of the 
mass-range set to which the data point belonged to. The weight we
assigned was simply proportional to the inverse of the size of
the set: $N/N_c$, where $N$ is the total number of planets and $N_c$
is the size of the set.  Table 1 details the three 
mass-range sets, their sizes and the corresponding weights. The final
expression for S is thus: 
$$ S(a_1,b_1,a_2,b_2) = \sum_{i=1}^N w_i \ \mathrm{min}
\Big\{ \frac{(y_i - a_1 x_i - b_1)^2}{a_1^2(\Delta x_i)^2 + (\Delta
y_i)^2} \ , \frac{(y_i - a_2 x_i - b_2)^2}{a_2^2(\Delta x_i)^2 +
(\Delta y_i)^2} \Big\} \ \ , 
$$ 
where $w_i$ is the weight of each point.

After optimizing $S$, we went on to obtain error estimates for the
four variables, using a Monte-Carlo resampling approach. We
randomly drew new data points from a Gaussian distribution.  The
expected values of the Gaussian distribution were the nominal
values of $x$ and $y$, and we used the error bars as the widths
(standard deviations) of the Gaussian distribution.  We repeated the
resampling procedure for $100,000$ such random realizations of the
data. The resulting random sample yielded the error estimates.

\section{Results}

Using the approach we outlined in the previous section, we obtained
estimates for the two slopes, and the breakpoint. We found the
breakpoint at a mass of $124 \pm 7 M_\oplus$, and a radius of $12.1
\pm 0.5 R_\oplus$. The resulting power laws of the two regimes
(based on the two slopes in the $x$-$y$ plane) are: $R \propto M^{0.55
\pm 0.02}$ for small planets, and $R \propto M^{0.01 \pm 0.02}$ for
large planets. The bottom panel of Figure 1 shows the derived
relation.
\par

It is interesting to note also according to our analysis, Saturn is "a small planet" (e.g., Chen \& Kipping 2017;  Weiss et al.\ 2013). 
Indeed, based on internal structure models, the heavy element fraction is Saturn is estimated to be between $\sim$ 20\% and 40\% (e.g., Guillot 2005). 
Thus, Saturn's mass is not very far from the transition point, and it is important to note that the transition mass at $\sim$120 $M_\oplus$ must be understood as a statistical quantity.  
As can be seen in Fig.~1, there is a region near the breakpoint in the fit at 120 $M_\oplus$ that could either be considered as the continuation of the regime where the radius increases with mass to even higher masses, or as an continuation of the high-mass regime (with approximately constant radius) to even lower masses. 
This transition regime approximately covers a mass range larger than the one derived in the analysis, somewhere between about 80 and 120 $M_\oplus$.  
Thus, according to the data, the actual transition occurs at the higher end of this mass range. Another point that should be taken into account is that the apparent transition is also affected by stellar irradiation, while Saturn experiences a much lower irradiation than most of the planets that were used in our statistical analysis. 
\par

Our results are in good agreement with previous studies.  The analysis
we used to obtain was simple and intuitive and did not rely on
subjective estimates.  The fact that the transition occurs at a
planetary mass larger than that of Saturn's supports the idea that the
change in the M-R relation for large planets is due to the dominating
composition -- in the case of massive planets, a mixture of hydrogen
and helium. The data suggest that for planets larger than $\sim 120
M_{\oplus}$, the planetary radius is determined by the equation of
state of these light elements (e.g., Zapolsky \& Salpeter 1969;
Fortney et al.\ 2007). The dominating H-He composition and the
compression due to the large mass also naturally explains the weak
dependence of the radius on mass for giant planets that consists of
mostly hydrogen and helium (e.g., Guillot 2005). Lower-mass planets
are less compressed and therefore, have a radius that increases in
mass. The relatively large spread of the low-mass planets around the
line suggests that in this mass regime, the planets can have various
compositions.

\subsection{Comparison with theoretical calculations}

In this section we briefly compare the observational data with
theoretical results from planet population syntheses based on the core
accretion paradigm (Mordasini et al.\ 2012). These calculations yield
the planetary bulk composition (solids and H/He) and the
post-formation entropy based on the planets' formation track. Here we
use two sets of cores (heavy-element) compositions: silicates and iron
or water. These two sets are chosen to assess the impact of various compositions of the solid core on the predicted radii of the synthetic planets. 
The first core is differentiated, and its composition is assumed to consists in mass of 1/3 iron (inner core) and 2/3 perovskite (outer core), similar to Earth and several low-mass extrasolar planets (e.g., Santos et al.\ 2015). The second composition corresponds to cores consisting exclusively of water ice. While pure water core are unlikely to exist, these cores represent the limiting case of low-density cores. In all cases, the modified polytropic equation of state is used to derive the core radius, taking into account the pressure exerted by the surrounding envelope (see Mordasini et al.\ 2012).  
The star is assumed to be $1 M_{\odot}$. Planets with
semi-major axes of 0.01 to 0.5 AU are included in order to have a
better comparison with the measurements. The formation model includes
the effect of type I and II orbital migration. During the evolutionary
phase, no mechanisms that can lead to inflation of the planetary radius (bloating) are included, whereas the effect of
atmospheric escape is considered as described in Jin et
al.\ (2014). 
The planetary opacity used in the formation models is the combination of the ISM opacities (Bell \& Lin 1994) reduced by a factor 0.003 plus the grain-free opacities of Freedman et al.\ (2014). 
The reduction factor was determined in Mordasini et al.\ (2014) by comparison with detailed simulations of the grain dynamics by Movshovitz et al.\ (2010). 
During the planetary evolution, we assume a grain-free opacity because grains are expected to grow and settle to deeper regions after gas accretion is terminated accretion stops (e.g., Movshovitz \& Podolak 2008). 
\par

The observations and the theoretical data are compared in Figure 3.
As can be seen from the figure, the general M-R relation is similar,
but there are also important differences. Both data sets show two
different regimes. In the low-mass regime, both the observational and
synthetic data show a large scatter in the M-R relation, which stems
in the synthetic population from different envelope-core mass ratios,
which in turn reflect different formation histories. For giant
planets, the simulated planets follow a narrow M-R relation, which is
clearly a consequence of neglecting bloating, assuming solar opacity,
and an internal structure consisting of a pure H/He envelope
surrounding a core made of pure heavy elements (i.e., a core+envelope
internal structure). This is in contrast to the observations that also
contain planets that have significantly larger radii, and probably
different compositions and/or internal structures. In addition, the
theoretical data correspond to a given age ($5 \times 10^9$ years),
while the observed population includes various ages. However, since most of the detected planets are observed around relatively old stars we do not expect a large impact on the goodness of fit to the observed M-R relation. 

In the giant planet regime, one sees that in the observed exoplanet population,
there are both giant planets with significantly larger, but also
smaller radii. The large radii can be attributed to bloating, while
the smaller planets suggest that there are some planets that contain
significantly higher amounts of heavy elements than in the synthetic
population. This could be the result of a more efficient accretion of
solids during formation, or giant impacts at later times. 
The effect of bloating on the population of small planets still needs to be studied in detailed although some work on this topic has already been presented (e.g., Lopez et al.\ 2012; Owen \& Wu 2013). At the moment, it is still unclear whether an inflation mechanism is required in order to explain some of the small exoplanets with very low mean densities, since the existence of an (H-He) atmosphere can significantly increase the planetary radius. In addition, unlike massive planets which are expected to be H-He dominated, small planets have a large spread of heavy-elements and various fractions of H-He. This introduces a degeneracy with inflation mechanisms for low-mass planets: an observed M-R relation can probably either be caused by the existence of a more massive H-He envelope without inflation, or alternatively by a physical mechanism that causes the planet to be large, i.e., inflation. A better understanding of inflation and atmospheric loss in small- 
  and intermediate- mass planets is clearly desirable.  
\par

Clearly, the two data sets should be compared only qualitatively. This
is because the observed planets have a variety of ages, atmospheric
opacities, and of course, possibly mixed compositions. As a result,
the partially strong and tight correlations in the theoretical M-R
should not be considered realistic, as they simply represent the
composition of pure ice/rock planets (in the case of the bare cores),
or the artificially narrow M-R relation of giant planets having all
the same atmospheric opacity and lacking bloating
mechanism. Nevertheless, there is a rather good agreement in terms of
the transition between "small" and "large" planets in the M-R diagram.

\section{Discussion and Conclusions}

Our analysis suggests that the transition between large and small
planets occurs at a mass (radius) of $124 \pm 7 M_{\oplus}$ ($12.1
\pm 0.5 R_{\oplus}$).  As expected, we establish two mass-radius
relations for exoplanets.  For low-mass planets, the radius is
increasing with increasing mass, and the M-R relation we derive is:
$R\propto M^{0.55\pm0.02}$, whereas for the large planets, the radius
is almost independent of the mass, and the M-R relation is $R \propto
M^{0.01\pm0.02}$.

Planetary mass and heavy element content almost exclusively determine
the radius of low-mass planets $< 124 M_{\oplus}$. The turnover point
at this mass is probably due to the characteristic boundary between
planets that are mostly gaseous (H-He dominated) and planets that
consist of varying compositions, and therefore, do not have a single
M-R relation.  When the planet mass reaches $> 124 M_{\oplus}$ the
relation is flattened and is even consistent with a small
negative slope, since we are approaching a slope of a compressed
hydrogen-helium-dominated planet.
\par 

This work identifies the transition point between small and large planets based on the M-R relation. 
This transition point is not the same as the one derived from studies of measured frequency of planets (occurrence rate), although the two might be linked. 
From standard planet formation models point of view, the transition from a heavy-element-dominated composition to a hydrogen-helium-dominated composition occurs at a mass where the core and envelope mass are similar (crossover mass). 
Statistical simulations of planet formation have shown (e.g., Mordasini et al.\ 2015) that this leads to a break in the planetary occurrence rate at about 30 $M_{\oplus}$, but the actual value can vary significantly depending on the assumed solid-surface density, opacity, accretion rates, etc. 
Thus, it is interesting to note that there are not many observed planets with masses between 30 and 120 $M_{\oplus}$ (see Fig.~1; see also Mayor et al.\ 2011). This may suggest that the two transitions are linked. Finding the link between the two transition points can reveal crucial information on planetary formation and characteristics, and we hope to address this topic in the future.  
\par

As mentioned earlier, thinking about planetary characterization in
terms of M-R relation is useful, but it should be noted that in
reality, there is a M-R-flux, or even
M-R-flux-time relation for planets. 
This is because the stellar flux and the time evolution are expected to affect the radius of the planet at a given time. 
These relations will
be better understood in the future when exoplanet detections will
include larger radial distances and various ages of stars, as expected
from the PLATO mission.

\subsection*{Acknowledgments} 
We thank the anonymous referee for valuable comments and suggestions. 
R.~H.~acknowledges support from the Israel Space Agency under grant 3-11485 and from the United States - Israel Binational Science Foundation (BSF) grant 2014112. 
C.~M.~acknowledges support from the Swiss National Science Foundation under grant BSSGI0$\_$155816 ``PlanetsInTime'. Parts of this work have been carried out within the frame of the National Centre for Competence in Research PlanetS supported by the SNSF.

\newpage
\section*{References}{}
Baglin, A., Auvergne, M., Boisnard, L., et al., 2006, 36th COSPAR Scientific Assembly, 36, 3749\\
Baraffe, I., Chabrier, G., Fortney, J. \& Sotin, C., 2014, in Protostars and Planets VI, Henrik Beuther, Ralf S. Klessen, Cornelis P. Dullemond, and Thomas Henning (eds.), University of Arizona Press, Tucson, 763\\
Bell, K.~R.~\& Lin, D.~N.~C., 1994, ApJ, 427, 987\\
Borucki, W.~J., Koch, D., Basri, G., et al., 2010, Science, 327, 977\\
Chen, J.~\& Kipping, D.~M., 2017, ApJ, 834, 17\\
Cumming, A., Butler, R.~P., Marcy, G., W.~et al., 2008, PASP, 120, 531\\
Durbin, J., 1954, Rev. Inst. Int. Stat., 22, 23\\
Estivill-Castro, V., 2002, SIGKDD explor., 4, 65\\
Fortney, J.~J., Marley, M.~S.~\& Barnes, J. W., 2007, ApJ, 659, 1661\\
Freedman, R.~S., Lustig-Yaeger, J., Fortney, J.~J., et al., 2014, ApJS, 214, 25\\
Golub, G., 1973, SIAM rev., 15, 318\\
Golub, G. \& Van Loan, C., 1980, SIAM J. Numer. Anal., 17, 883\\
Guillot, T., 2005, Annu. Rev. Earth Planet. Sci., 33\\
Hatzes, A.~P.~\& Rauer, H., 2015, ApJL, 810, L25 \\
Hinkley, D. V., 1969, Biometrika, 56, 495\\
Howard, A., Marcy, G.~W., Johnson, J.~A.~, et al., 2010, Science,  330, 653\\
Howard, A.~W.,  Marcy, G.~W., Bryson, S.~T.~et al., 2012, ApJS, 201, 15\\
Jin, S., Mordasini, C., Parmentier, V., van Boekel, R., Henning, T., Ji, J.,  2014, ApJ, 795, 65\\
Lopez, E.~D.,  Fortney, J.~J.~\& Miller, N., 2012, ApJ, 761, 59\\
Markovsky, I., Sima, D. M. \& Van Huffel, S., 2010, WIREs Comp. 
Stats. 2, 212\\
Marcy, G.~W.,  Weiss, L.~M., Petigura, E.~A., Isaacson, H., Howard, A.~W.~\& Buchhave, L.~A., 2014, Proc. Natl. Acad. Sci. USA, 111, 12655\\
Mayor, M., Marmier, M., Lovis, C.~et al., 2011, A\&A, submitted. arXiv:1109.2497\\
Miller, N.~\& Fortney, J.~J., 2011, ApJ, 736, L29 \\
Mordasini, C., Alibert, Y., Georgy, C., et al.,~2012, A\&A, 547, A112\\
Mordasini, C., Klahr, H., Alibert, Y., Miller, N.~\& Henning, T., 2014, A\&A, 566, A141\\
Mordasini, C., Molli{\'e}re, P., Dittkrist, K.~M., Jin, S.~\& Alibert, Y., 2015, Int. J. Astrobiol., 14, 201\\
Movshovitz \& Podolak, 2008, Icarus, 194, 368\\
Movshovitz, N., Bodenheimer, P., Podolak, M.~\& Lissauer, J.~J., 2010. Icarus, 209, 616\\
Nelder, J.~A.~\& Mead, R., 1965, Comput. J., 7, 308\\
Nesvorn\'{y}, D.~\& Morbidelli, A., 2008, ApJ, 688, 636\\
Owen, J.~E.~\& Wu, Y., 2013, ApJ, 775, 105\\
Santos, N. C., Adibekyan, V., Mordasini, C.~et al., 2015, A\&A, 580, L13\\
Weidenschilling, S. J., 1977, ApSS,  51, 153\\
Weiss, L.~M., Marcy, G.~W., Rowe, J.~F.~et al., 2013, ApJ, 768, 14\\
Weiss, L.~M.~\& Marcy, G.~W., 2014, ApJ, 783, L6\\
Wooldridge, J. M., 1999, Econometrica, 67, 1385\\
Zapolsky, H.~S.~\& Salpeter, E.~E., 1969, ApJ, 158, 809 

\newpage
\begin{table}[h!]
\center{
{\renewcommand{\arraystretch}{1.7}
\caption{\label{clusters} 
The details of the three mass-range sets, and the resulting weights used in the analysis.}
\begin{tabular}{ | c | c | c | c |}
\hline
\hline
 & {\bf Group A} & {\bf Group B} & {\bf Group C} \\
 \hline
$M$ &  $M<69 M_{\oplus}$ & $69 M_{\oplus} < M < 1660 M_{\oplus}$ & $M>1660 M_{\oplus}$ \\
\hline
$N_K$ & $N_A = 54$ & $N_B = 207$ & $N_C = 13$\\
\hline
$W_K$ & $W_A = 5.0741$ & $W_B = 1.3237$ & $W_C = 21.0769$ \\
\hline
\hline
\end{tabular} 
}
}
\end{table}

\begin{table}[h!]
\center{
{\renewcommand{\arraystretch}{1.7}
\caption{The M-R relations derived for small and large planets.}          
\label{exoplanets}      
\centering          
\begin{tabular}{| c | c | c | c | l l l }     
\hline\hline       
  & {\bf Weiss et al. } & {\bf Chen \& Kipping} & {\bf This work} \\
    & {\bf 2013} & {\bf  2017} &  \\
 \hline
{\bf Small planets} & $R \propto M^{0.53\pm0.05}$ & $R \propto M^{0.59^{+0.04}_{-0.03}}$ & $R \propto M^{0.55\pm0.02}$ \\
\hline
{\bf Large planets} & $R \propto M^{-0.04\pm0.01}$ & $R \propto M^{-0.04\pm0.02}$ & $R \propto M^{0.01\pm0.02}$ \\
\hline 
\hline 
\end{tabular}
}
}
\end{table}

\newpage
   \begin{figure}
\centerline{\includegraphics[angle=0, width=12cm]{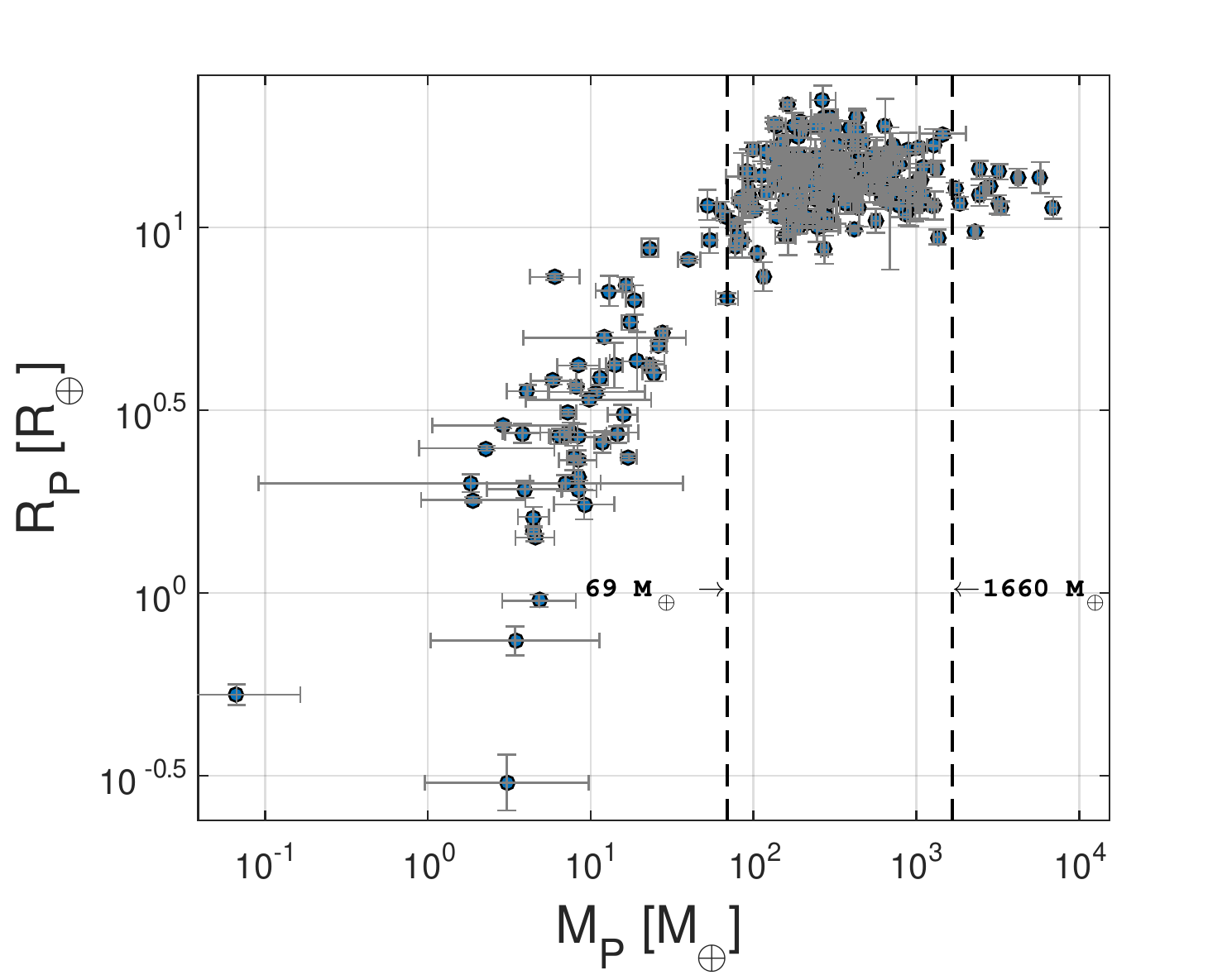}}
\vspace{-0.1cm}
\centerline{\includegraphics[angle=0, width=12cm]{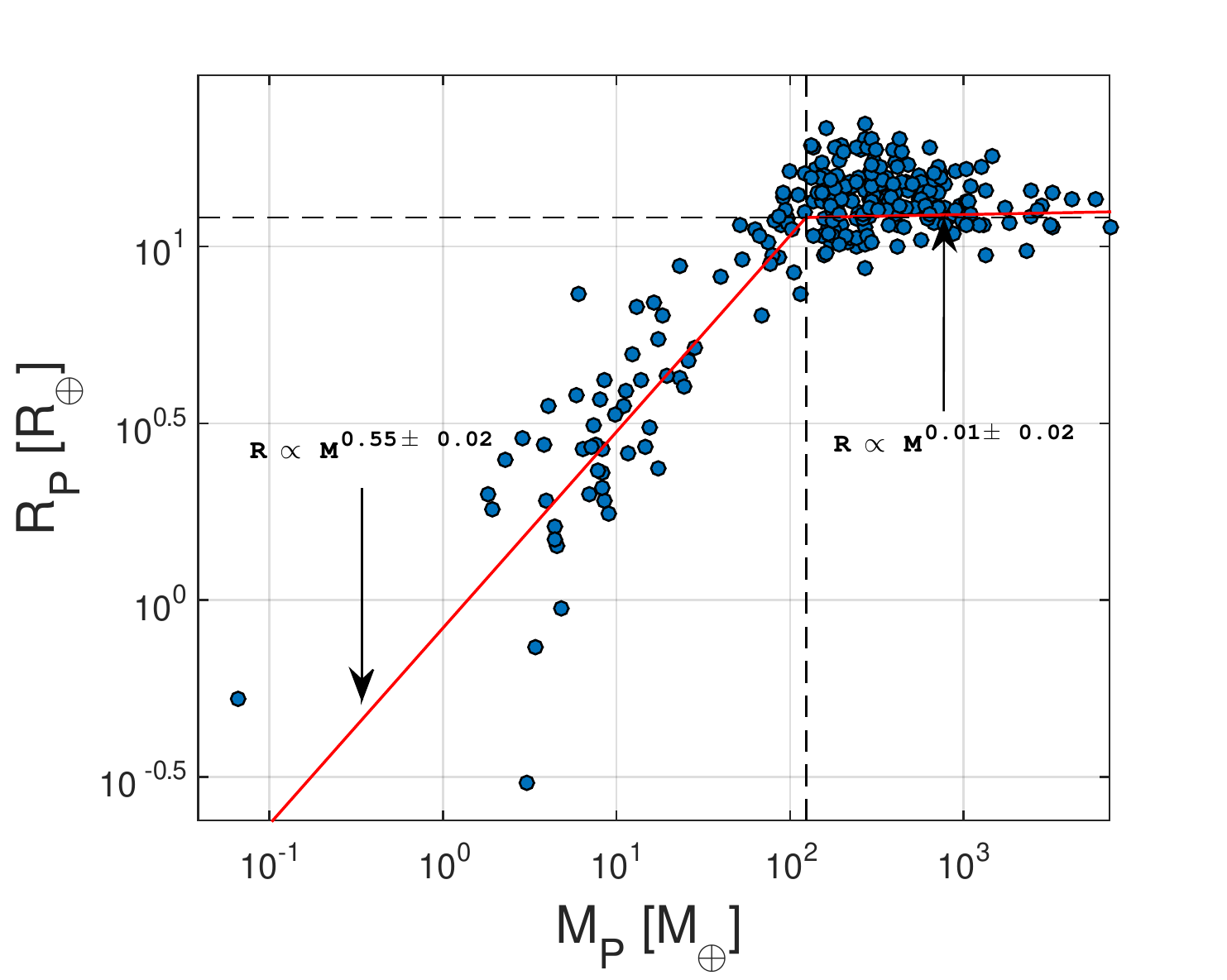}}
\vspace{-0.1cm}
\caption{{\bf Top:} The M-R relation of the exoplanets considered in the analysis.  The dashed lines identify the three different regimes we consider for the weighting (see text) {\bf Bottom:} The M-R relation and the derived  best-fit curves, and M-R relations.}
\label{MR}
   \end{figure}
%

\newpage
\begin{figure}
\centerline{\includegraphics[angle=0, width=15cm]{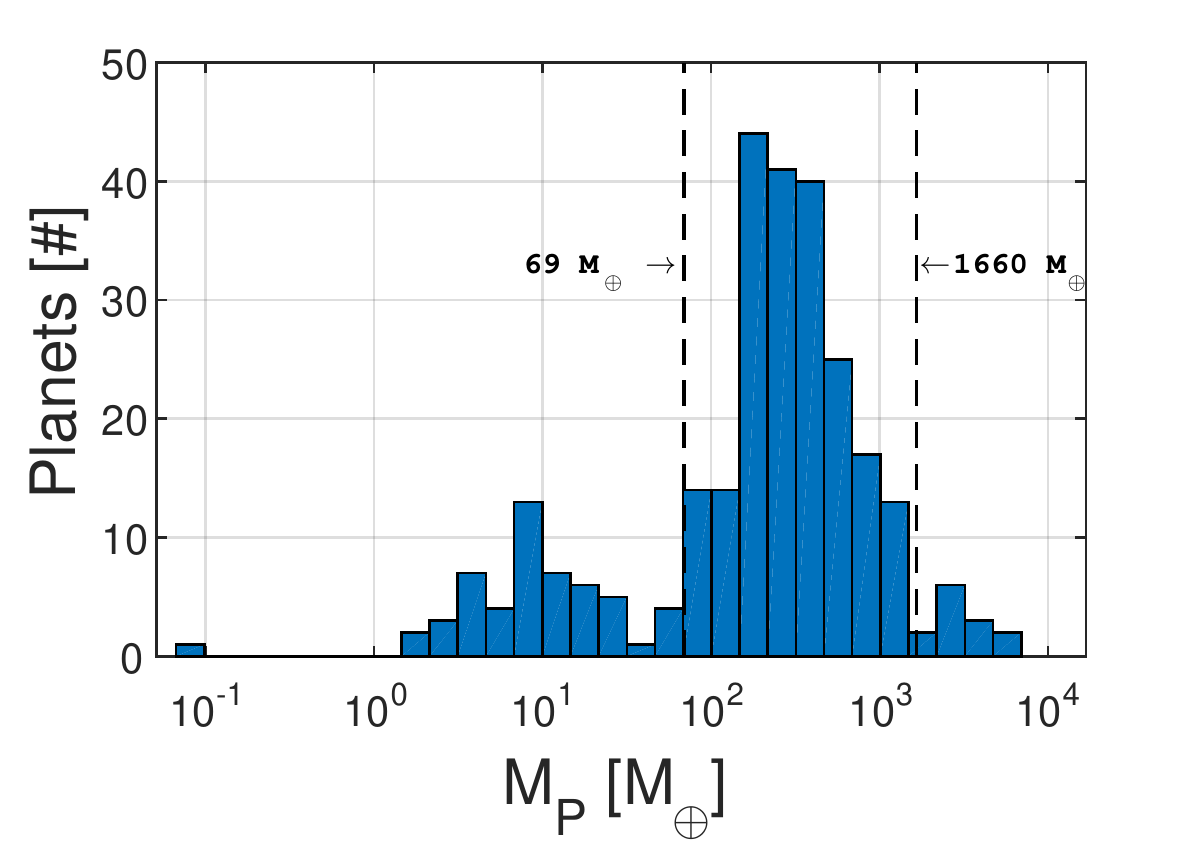}}
\caption{Histogram of planetary mass, showing clearly the three 
 empirical mass ranges. The division into three intervals
 was performed in order to improve the quality of the statistical
 analysis (see text for details).}
\label{mass_hist}
\end{figure}

\newpage
\begin{figure}
\centerline{\includegraphics[angle=0, width=15cm]{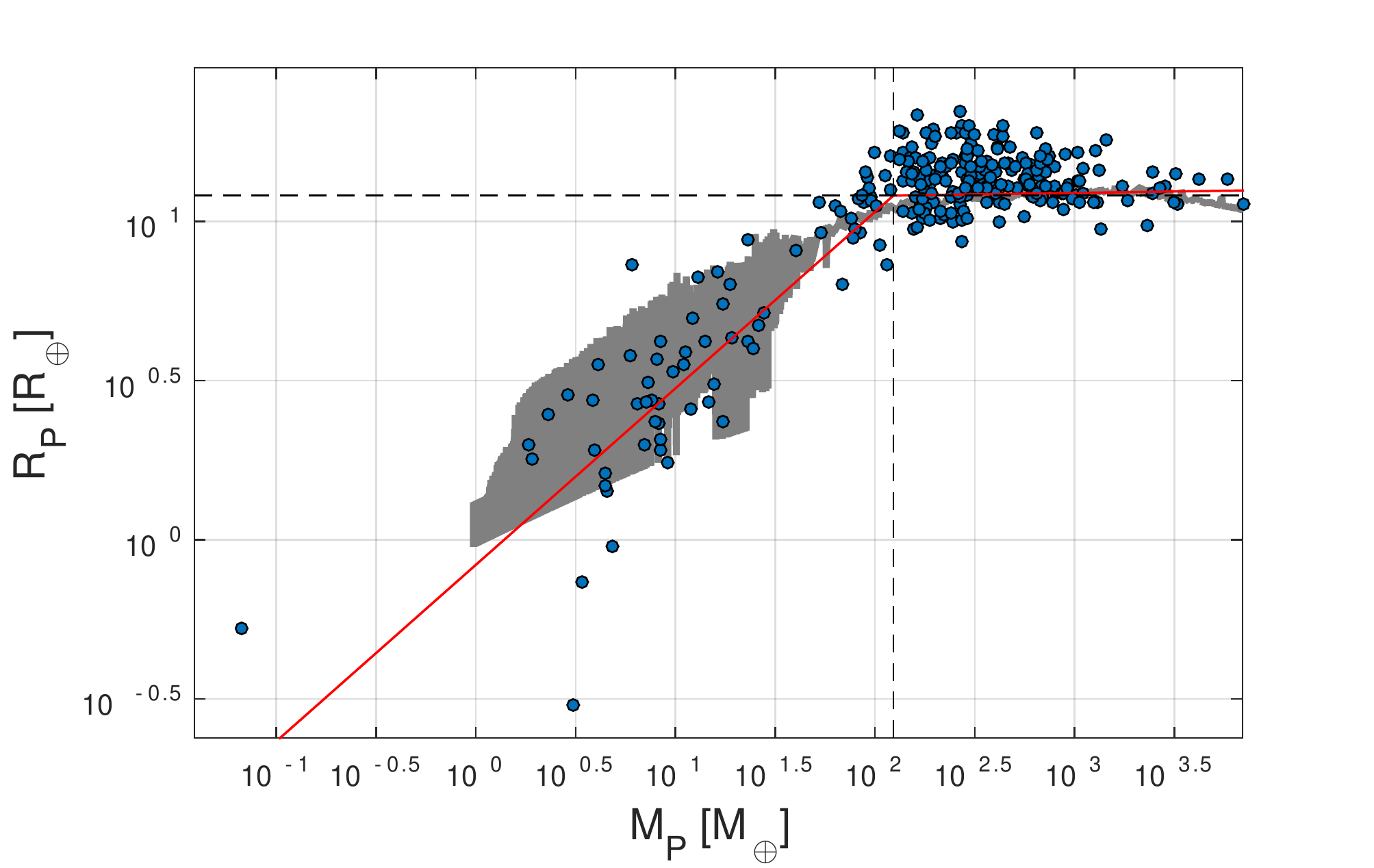}}
\caption{The M-R relation: observations vs.~theoretical data. The circles correspond to the observations while the shaded area represents the results from planet population synthesis models.}
\label{chi}
\end{figure}

\end{document}